\documentclass[Lau,binding=0.6cm]{sapthesis}
\usepackage{float}
\usepackage[utf8]{inputenc}
\usepackage{listings} 
\usepackage[dvipsnames]{xcolor}   
\usepackage{hyperref}
\usepackage{graphicx}
\usepackage{subfig}

\graphicspath{ {images/} }

\hypersetup{
    colorlinks,
    citecolor=black,
    filecolor=black,
    linkcolor=black,
    urlcolor=black
}

\lstnewenvironment{cpp_code}[1][]{
    \lstset{
        language=C++,
        numbers=left,
        stepnumber=1,
        breaklines=true,
        basicstyle=\linespread{1.0}\ttfamily,
        keywordstyle=\color{blue}\ttfamily,
        stringstyle=\color{red}\ttfamily,
        commentstyle=\color{OliveGreen}\ttfamily,
        morecomment=[l][\color{magenta}]{\#}
        columns=fullflexible,
        postbreak=\mbox{\textcolor{red}{$\hookrightarrow$}\space},
        escapeinside={(*@}{@*)},
        showstringspaces=false
    }
    \lstdefinestyle{nonumbers}
    {numbers=none}
}{}

\lstnewenvironment{py_code}[1][]{
    \lstset{
        language=python,
        stepnumber=1,
        breaklines=true,
        basicstyle=\linespread{1.0}\ttfamily,
        keywordstyle=\color{blue}\ttfamily,
        stringstyle=\color{red}\ttfamily,
        commentstyle=\color{OliveGreen}\ttfamily,
        morecomment=[l][\color{magenta}]{\#}
        columns=fullflexible,
        otherkeywords={self},
        postbreak=\mbox{\textcolor{red}{$\hookrightarrow$}\space},
        escapeinside={(*@}{@*)},
        showstringspaces=false,
    }
    \lstdefinestyle{nonumbers}
    {numbers=none}
}{}

\lstnewenvironment{code}[1][]{
    \lstset{
        language=bash,
        breaklines=true,
        basicstyle=\linespread{1.0}\ttfamily,
        keywordstyle=\color{blue}\ttfamily,
        stringstyle=\color{red}\ttfamily,
        commentstyle=\color{OliveGreen}\ttfamily,
        morecomment=[l][\color{magenta}]{\#}
        columns=fullflexible,
        postbreak=\mbox{\textcolor{red}{$\hookrightarrow$}\space},
        escapeinside={(*@}{@*)},
    }
    \lstdefinestyle{nonumbers}
    {numbers=none}
}{}

\title{Symbolic Execution and \\Debugging Synchronization}
\author{Andrea Fioraldi}
\IDnumber{1692419}
\course{Computer and System Engineering}
\courseorganizer{Faculty of Information Engineering, Informatics, and Statistics}
\AcademicYear{2017/2018}
\copyyear{2018}
\advisor{Prof. Camil Demetrescu}
\coadvisor{Dr. Emilio Coppa}
\coadvisor{Dr. Daniele Cono D'Elia}
\authoremail{fioraldi.1692419@studenti.uniroma1.it}

\versiondate{\today}

\begin{document}

\frontmatter

\maketitle

\clearpage

\vspace*{\stretch{1}}
\textit{We, hackers, are the dragons of the digital age.}

Order of the Overflow
\vspace*{\stretch{1}}

\clearpage

\tableofcontents

\clearpage

\chapter{Introduction}

Binary analysis is one of the most important problems in computer security.
Despite the rise of web applications and interpreted languages like Python, the are several reasons that justify this importance, as described in \cite{shoshitaishvili2016state}.
Nowadays, operating systems are still based on compiled languages and also interpreted languages are often compiled down during execution by a Just In Time compiler. In addition, traditional low-level languages like C are reborn thanks to the rise of Internet Of Things and the associated resource-constrained devices.

A wide family of manual and automatic techniques has been developed over the years to analyze this kind of programs.
These techniques are divided into static and dynamic analyses depending on the need to execute the program or not.

Despite the continuous innovation on the front of the automatic analysis, the human part remains essential.
A reverse engineer is a person who tries to understand what a binary program does and how.
This process usually involves reading and understanding the disassembled code and its effects during execution with the aid of a debugger.

One of the most used automatic techniques nowadays is {\em Symbolic Execution}.
The idea came about in the '70s \cite{King} but only recently its applications became relevant in computer security.
Symbolic execution is used for different tasks, varying from deobfuscation to vulnerability detection and automatic exploit generation.

Symbolic execution has been traditionally considered a static analysis technique since in its original embodiment the application code is not executed concretely by the CPU. However, a more recent twist of symbolic execution, known as Dynamic Symbolic Execution (DSE) \cite{DART}, uses a concrete execution to drive the symbolic exploration and thus can be classified as a dynamic analysis technique.

In the last years, symbolic execution becomes also a first-class technique used by anyone who deals with manual reverse engineering.


Execute symbolically an entire complex software (like a web server for example) is a huge task for a machine, even for a supercomputer so the analyst often uses it in a surgical manner on small pieces of code.
For instance, a reverse engineer often needs to reverse custom encryption functions or obfuscated code during the dynamic analysis.

In this thesis, we introduce the idea of combining symbolic execution with dynamic analysis for reverse engineering.

Differently from DSE, we devise an approach where the reverse engineer can use a debugger to drive and inspect a concrete execution engine of the application code and then, when needed, transfer the execution into a symbolic executor in order to automatically identify the input values required to reach a target point in the code. After that, the user can also transfer back the correct input values found with symbolic execution in order to continue the debugging.

The synchronization between a debugger and a symbolic executor can enhance manual dynamic analysis and allow a reverser to easily solve small portions of code without leaving the debugger.

We implemented a synchronization mechanism on top of the binary analysis framework {\em angr}, allowing for transferring the state of the debugged process to the angr environment and back.

The backend library is debugger agnostic and can be extended to work with various frontends.
We implemented a frontend for the {\em IDA Pro} debugger and one for the {\em GNU Debugger}, which are both widely popular among reverse engineers.

\subsubsection{Structure of the Thesis}

Chapter 1 describes what is symbolic execution and its limitations. Chapter 2 presents our technique and discuss advantages and limitations. Chapter 3 is about the implementation of the technique in a library based on angr, {\em AngrDBG}, and its frontends, {\em AngrGDB} and {\em IDAngr}. Chapter4 is a usage example of IDAngr to show how the tool works.

\mainmatter

\chapter{Symbolic Execution}

In this chapter, we will present the main ideas behind {\em symbolic execution}, discussing its limitations and why the current approaches may be not adequate for reverse engineering of real-world software.

\section{Overview}

When a program is executed on a machine the CPU follows a single flow of instructions.
The control flow path can be different for each concrete execution of the program and only a single path can be explored at a time.
Hence, it is not possible to evaluate all the possible behaviors of a program when analyzing a single execution.

{\em Symbolic execution} is a program analysis technique able to explore simultaneously different paths in a program, evaluating how different program inputs may affect the execution flow.

While a concrete execution evaluates the program execution based on the concrete values associated with the program inputs, symbolic execution evaluates the program execution using {\em symbolic} values associated with them.

A symbolic value may initially assume any value within the data type domain. When evaluating the program statements, symbolic execution constructs expressions over the symbolic inputs to describe the manipulation performed on the program inputs. When a branch is encountered during the exploration, symbolic execution checks the satisfiability of the branch condition using an SMT \footnote{Satisfability Modulo Theories} solver.

If both branches are feasible, the exploration is forked, adding to each execution path a new constraint that restricts the values allowed for the program inputs in that execution path. The set of constraints collected during the exploration of a path are typically referred to with the term {\em path constraints}. Additionally, the {\em symbolic store} is used to keep track of the mapping between program variables and symbolic or concrete expressions constructed during the exploration

\subsection{Example}

To better explain how symbolic execution works, let us consider the following piece of code:

\begin{cpp_code}
int foo(int a, int b)
{
    int c = 77;

    if(a + b == 42) {
        c = c - b;
    }
    else {
        c = a - c;
    }
    if(c == 38) {
        puts("Well done.");
    }
    else {
        puts("Try again.");
    }
}
\end{cpp_code}

The goal is to find some valid values for the input variables a and b to reach line 12.

Before starting the evaluation with symbolic execution, two symbolic values, $\delta_{\mathrm{a}}$ and $\delta_{\mathrm{b}}$, are associated with the program inputs a and b, respectively.

In the beginning, we have only a path, with the variable $c$ associated with a concrete value, 77.

When we reach the first if statement two different paths are generated with the following formulas and symbolic storages:
\begin{itemize}
\item \verb|[Path #1]| Formula: $\delta_{\mathrm{a}} + \delta_{\mathrm{b}} = 42$, Storage: $c = 77$
\item \verb|[Path #2]| Formula: $\neg(\delta_{\mathrm{a}} + \delta_{\mathrm{b}} = 42)$, Storage: $c = 77$
\end{itemize}

In the next step for Path 1 the symbolic storage changes to $c = 77 - \delta_{\mathrm{b}}$.
Similarly, the Path 2 storage changes to $c = \delta_{\mathrm{a}} - 77$.

So Path 1 and 2 reach both the second if statement with different constraints sets.
We want to reach the branch associated to the condition $c == 38$.

There are four different possible paths:
\begin{itemize}
\item \verb|[Path #1.1]| Formula: $\delta_{\mathrm{a}} + \delta_{\mathrm{b}} = 42 \land c = 38$, Storage: $c = 77 - \delta_{\mathrm{b}}$
\item \verb|[Path #1.2]| Formula: $\delta_{\mathrm{a}} + \delta_{\mathrm{b}} = 42 \land c \neq 38$, Storage: $c = 77 - \delta_{\mathrm{b}}$
\item \verb|[Path #2.1]| Formula: $\neg(\delta_{\mathrm{a}} + \delta_{\mathrm{b}} = 42) \land c = 38$, Storage: $c = \delta_{\mathrm{a}} - 77$
\item \verb|[Path #2.2]| Formula: $\neg(\delta_{\mathrm{a}} + \delta_{\mathrm{b}} = 42) \land c \neq 38$, Storage: $c = \delta_{\mathrm{a}} - 77$
\end{itemize}

The interesting paths are 1.1 and 2.1 because they reached our target, the \verb|puts("Well done.");| statement. By solving their collected constraints we can get some concrete values for the symbolic inputs that allow the program to reach the target branch during a concrete execution.

The final path constraints for Path 1.1 are $\delta_{\mathrm{a}} + \delta_{\mathrm{b}} = 42 \land 77 - \delta_{\mathrm{b}} = 38$ and solving it with a SMT solver gives this result: $\delta_{\mathrm{a}} = 3$, $\delta_{\mathrm{b}} = 39$.
In fact $3 + 39 = 42 \land 77 - 39 = 38$.

Since an assignment for the program input able to reach the target has been found, the symbolic execution can be terminated. If a user may want to get multiple assignments to reach the target, Path 2.1 can be evaluated to obtain another set of concrete values for the program inputs from the SMT solver.

They final path constraints for 2.1 is $\neg(\delta_{\mathrm{a}} + \delta_{\mathrm{b}} = 42) \land \delta_{\mathrm{a}} - 77 = 38$ and it evaluates to:  $\delta_{\mathrm{a}} = 115$, $\delta_{\mathrm{b}} = 40$.

This is an example of what a symbolic executor does.

\section{Dynamic Symbolic Execution}

A main limitation of symbolic execution is exploring paths with very complex constraints.
The time spent on solving these constraints in order to know if a path can be reached is critical in terms of engine's efficiency.

A game-changing technique to address this problem is {\em Dynamic Symbolic Execution (DSE)}, the combination of symbolic and concrete execution, introduced by \cite{DART}. The idea is to drive symbolic execution along recorded concrete execution paths and so avoid the call to the solver to know if a path is unreachable. This is possible due to binary instrumentation and also permits a simplified implementation of the symbolic interpreter.


Dynamic symbolic execution is often implemented with the possibility of selective concretization of some values in order to simplify constraints.

\subsection{Algorithm}

The high-level steps of the DSE process are:

\begin{enumerate}
\item Choose input variables and set them as symbolic;
\item Instrument the program to trace all events regarding the input variables;
\item Choose a random input and execute the program;
\item Execute the symbolic execution engine on the traced path and collect all constraints;
\item Negate the last path condition in order to visit a new path. If there is not a path condition left then terminate the exploration;
\item Invoke the solver to generate a new input. If the condition is unsatisfiable go to step 5;
\item Execute the program with the generated input and go to step 4;
\end{enumerate}

\subsection{Limitations}
\label{dynamic_limit}

Dynamic symbolic execution cannot work with not deterministic programs as the executor cannot analyze which path is executed with an input variation.
Another issue is when a code cannot be tracked by the engine. A side effect in that code may cause a wrong input generation that leads to execute an unintended path.

\section{Main challenges in symbolic execution}

As described before symbolic execution can, in theory, explore all possible paths in a binary. However, this is not always possible since symbolic execution may not be able in practice to analyze complex programs.
As described in \cite{SurveySymExec-CSUR18}, several problems make that hard:

\begin{enumerate}
\item How can a symbolic executor manage the dereferencing of a symbolic pointer?
\item How to deal with programs with a large number of branches (e.g., loops) that can easily make the number of paths exponential?
\item How model interactions with the environment?
\item How to solve non-linear constraints in a reasonable time?
\end{enumerate}

Some of these challenges may be addressed when using symbolic execution within a specific application context. However, the assumptions made to mitigate these issues may lead symbolic execution to miss some interesting paths or to follow unrealizable program paths.

In the remainder of this section, we discuss some of these problems in more details and present few approaches that have been proposed in the literature to mitigate them.

\subsection{Handling Symbolic Memory}

In the most general symbolic execution approach, a memory address may be symbolic and thus reference in the worst case the whole content of the memory.
This leads to an ambiguity for the dereferencing operation.
For example, consider a concrete array \verb|A| and a symbolic index \verb|i|. Since the index is symbolic, the expression \verb|A[i]| may actually refer to any element in memory. Although {\em symbolic} accesses are not rare in real-world programs, how to deal with these kinds of accesses remains an open problem.

There are two fundamental methods to handle this \cite{King}, even if neither is a solution that works at scale:

\begin{itemize}
\item {\em State Forking}: for each memory operation with symbolic addresses the state is forked considering all possible derived states
\item {\em If-Then-Else formulas}: this method exploits the capability of some constraint solvers of handling conditional operators in logical formulas, so the pointer is kept in the symbolic storage updating the path's boolean formula with the conditional operator avoiding forking.
\end{itemize}

However, when a pointer value can be in a very large range most engines do an address concretization, sometimes to NULL or to a newly allocated object.

Since concretization may lead a symbolic executor to miss many interesting paths, one approach proposed in literature \cite{mayem} is the {\em partial} memory model, where write accesses are always concretized, while read accesses are not concretized but treated symbolically when the symbolic address ranges a limited memory area (e.g., up to 1024 bytes).

\subsubsection{A note on loops}

When the number of iterations of a loop depends on the program input this can be a problem for the symbolic executor. In DSE if the engine follows exactly the same path of the recorded concrete execution the number of iterations of a loop is the number of iterations associated to the generated input and not all the possible iterations associated to the symbolic input.

The {\em LESE} \cite{Saxena:EECS-2009-34} technique was developed to bypass this problem. It introduces a symbolic trip-count variable for each loop that represents the number of iterations and mantains a relationship between the symbolic variables and the trip counts. A symbolic value thus has not only a data relationship with the input but also with the loop-dependent effects.

\subsection{Path Explosion}

When the number of forked states is exponential because of the number of branches symbolic execution can become pathologically slow.
Loops and calls are the main sources of the branch increase. Each loop iteration adds a conditional statement and if the loop depends on some symbolic values the number of branches can be infinite.

A first valid technique to reduce the number of explored states is to remove unsatisfiable paths as soon as possible. However, calling the solver too often can degrade performance, as we will discuss in \ref{constr_solv}.

Another valid technique to reduce forking is state merging. It consists in merging different paths into a single state with a formula that is the disjunction of the merged paths formulas. The disjunction formula makes use of the If-Then-Else conditional expressions.

There several further possible solutions and some are based on the under-approximation of the number of states to explore:

\begin{itemize}
\item Set a precondition on the input to reduce the number of explored states;
\item Bound loops to a limited number of iterations
\item Use path similarity to discard paths that cannot lead to new findings. A popular technique is interpolation, a method to decide if a formula is related to an undesired property and so discard the exploration of the associated path;
\item Record an execution summary of loop bodies and functions calls so when they are traversed again the symbolic executor can use previous results;
\end{itemize}

\subsection{Interaction With Environment}

Early implementations of symbolic execution were unable to symbolic reason on the interactions between a program and the operating system. In particular, a common approach was to concretize the arguments of library or system APIs and run concretely the interaction. However, using these approach, only the return value was taken into account, ignoring any side effects possibly performed by the operating system on the process memory. Additionally, concrete evaluation of interaction often resulted in missing interesting execution paths.

To overcome the inaccuracy resulting from concretization during environment interactions, a common approach is to create an abstract model that handles the interaction and provides a simulation of the environment (e.g a symbolic filesystem in order to support operations on files).
Models are implemented typically at the system call level because writing a model for all possible external functions that a program can use is unpracticable.
This leads to the symbolic exploration of the library code.

However, some engines combine system calls models with external functions models (typically the standard library API) and only when such models are missing the exploration of library code is performed. 

\subsection{Constraint Solving}
\label{constr_solv}

The {\em boolean satisfiability problem (SAT)} is an NP-complete problem. {\em Satisfiability modulo theories (SMT)} are used to generalize SAT with arithmetic operations and arrays.
Symbolic execution engines, as described before, use SMT solvers to evaluate logical constraints.

To make symbolic execution scalable the constraints solving process must be optimized.
There are some valid approaches:

\begin{itemize}
\item Reduce the size and complexity of the generated expressions;
\item Reuse solutions from previous (similar) queries;
\item Lazy evaluation: on a branch, the symbolic executor takes both paths adding a lazy constraint to the formula. These constraints are evaluated only when the path reaches a target and the path is; discarded if it is unreachable in a concrete execution. This increments the number of paths but also reduces the solution space of the solver thanks to the constraints added after the lazy constraint;
\item Concretization of not solvable expressions (e.g. complex arithmetics) with random values;
\end{itemize}

\section{Discussion}

Symbolic execution is a very large matter and we presented only a basic description of the most common issues and solutions. To understand deeply the state of the art the reader can refer to \cite{SurveySymExec-CSUR18}.

In the following chapters we will focus on one application of symbolic execution: help an analyst to understand the behavior of a program.
A real example is a malware analyst that wants to reconstruct the protocol of a command-and-control malware. The analyst can understand which command activates a specified part of the malware marking the input from the socket as symbolic and executing symbolically the malicious program.

The alternation of concrete execution and symbolic execution with the interaction of the analyst is a novel contribution in the symbolic execution field.

\chapter{Technique}

In this chapter, we will discuss the advantages and the limitations of combining symbolic execution with a debugger and, more in general, with any forms of concrete execution.

\section{Debuggers}

A {\em debugger}, according to \cite{PMA}, is a piece of software or hardware used to test or examine the execution of another program.
It is mainly designed to allow a developer to examine and control the internal state and the execution flow of a program.

Inspecting and also manipulating the execution context (the memory and the registers) of the program at any time is a fundamental operation during the reverse engineering process.

There are two types of debuggers from a user point of view:
\begin{itemize}
\item source-level, that allows the user to inspect the program state in terms of the source code when available;
\item low-level, with which the user operates only on disassembly;
\end{itemize}

Reverse engineering a software is a process that does not involve often the source code so low-level debuggers are the most used.

The main features of a debugger are the following:

\begin{itemize}
\item Single stepping: execute an instruction and return the control to the debugger;
\item Set a breakpoint: mark a place in the code or define a condition for which the execution is stopped so that the user can inspect the program state;
\item Modify execution: for example, change the value of a memory location to understand how a function works with a determinate input;
\item Hook exceptions: return the control to the debugger when an error like SEGFAULT occurs;
\end{itemize}

\section{Motivation}

The work of a reverse engineer is not simply reading assembly, often the code is hidden or very difficult to read and only with an automatic tool it can be understood.
A reverser makes an heavy usage of the debugger and in many cases the difficult code is revealed only during execution, the simplest example is a packed executable.

In a lot of malware samples, a target point in the program can be reached only under complex conditions, like in a malware with evasion techniques, and the program state can be very complex.

If we start symbolic execution from the entry point of the program the exploration may never reach the target point due to the limitations exposed in the previous chapter.
Even if that point is reached the path constraints and the symbolic expressions can be very complex and the exploration of the interesting code that follows the target point can be very expensive due to the time spent by the SMT solver to verify these complex expressions.

On the other hand, we would like to start the exploration directly from the beginning of the interesting code to avoid these issues. To this end we have to create a state from which to start the symbolic execution in a consistent way, setting the variables in the symbolic executor environment that are required for the correct behavior of the code that must be explored.
When done manually, this task can be time-consuming and error-prone.

However, the initial state for the symbolic execution may be created automatically by importing the concrete state from a concrete execution.

Since we do not know a priori which variables are involved during the execution of the code we need to take the entire concrete state and transfer it to the symbolic executor.

In this thesis, we propose to use a debugger to make easy for an analyst to reach the useful point.

This is the essence of our technique, transfer (even complex) states from a debugger to a symbolic executor and write back the results in order to synchronize both states. This makes possible for the reverser to continue the debugging after the exploration.

This technique can be very effective if there is the need to do symbolic execution of small pieces of code often during the debugging, so you can manually control the execution and solve difficult code in a surgical way.

Obviously, this approach is not limited to debuggers but can be extended to transfer the state from memory dumps, emulators or from a concrete execution controlled by a DBI framework\footnote{Dynamic Binary Instrumentation \cite{Net:phd2004}: a dynamic analysis technique that inserts analysis code at run-time in the "instrumented" program.} like Intel Pin \cite{Pin} or Frida \cite{Frida}.

\section{Description}
\label{descr}

If we try to imagine the simplest implementation of state transfer from a paused debugger to a symbolic executor it will be a simple copy of registers and memory.
However, this approach comes with severe disadvantages.

The address space of a process may be large thus copying all the memory is not practicable because of the slowness of the process. The memory must be requested to the debugger lazily.


Before the state transfer, the analyst must set the symbolic variables and its constraints. The constraints are required in order to force the symbolic executor to explore only the paths that can be executed starting from the current debugger state.
These constraints must not conflict with the required constraints that allow reaching of the concrete state that we are going to transfer. Since there is no previous symbolic exploration that collects these constraints this process requires an accurate analysis and a correct reverse engineering of the code executed before starting the state transfer. An error during this passage may compromise exploration and lead to unfeasible executions.

For example if we must set as symbolic a string read during the debugging with \verb|scanf("%s"...)| and we do not set also the constraints that it cannot contain whitespace characters, after the exploration the concretized string can contain whitespaces making it an input that can never be entered by the user.

In addition, a program state is not only registers and memory but also how the process interacted/must interact with the environment.
Environment synchronization is a challenging problem.

To do this we must build a mechanism to create a symbolic executor environment as similar as possible to the concrete environment. An example is synchronizing the position of the cursor associated with the opened file descriptors. In this thesis, we propose to interact with the environment saving the context of the concrete process, executing syscalls concretely and retrieve information from the operating system and then restoring the context just before the state transfer.

The reverse process, transfer the state found after the exploration to the debugger is not trivial. All the modification to the environment must be tracked and reproduced on the concrete environment, so the most reasonable way is to concretize the variables that we chose as symbolic input before the symbolic exploration, inject them in the debugger and run the paused process replicating the execution of the path found during the symbolic exploration.

\begin{figure}[H]
  \caption{High-level flow of the technique.}
  \centering
  \includegraphics[width=1\textwidth]{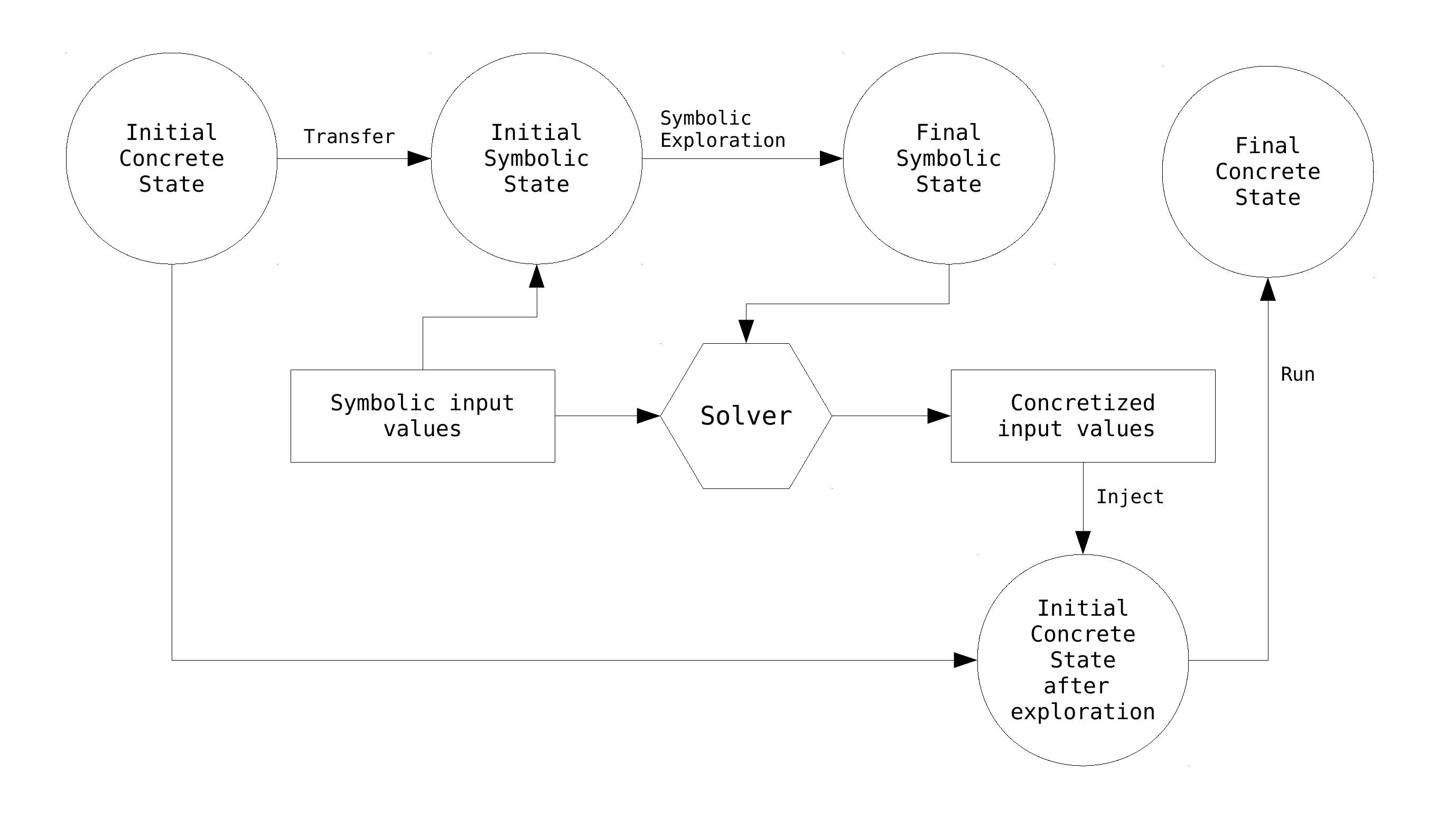}
  \label{fig:g}
\end{figure}

\subsection{Memory Transfer Strategies}

During the development of the technique we identified three different strategies for handling the retrieving of the memory from the concrete process:

\begin{enumerate}  
\item Transfer the entire process memory during state creation;
\item Lazy transfer of pages when requested;
\item Lazy transfer of bytes when requested;
\end{enumerate}

The first is a bad choice, as motivated in section \ref{descr}, since the process memory is huge and a state transfer can consume too much time and memory.
The other two are reasonable strategies and in this thesis, we have chosen the last one after a preliminary experimental evaluation that showed how moving even single pages in place of just a few bytes can add a significant overhead.





\section{Limitations}

A limitation of this approach is handling the process state in the operating system. In this section, we consider only the GNU/Linux operating system but the exposed concepts are the same for all operating systems.

A very important part of the environment is the metadata associated with a process in the operating system: the {\em Process Control Block}.

This record cannot be accessed from userspace and a synchronization tool working at user level, as in our design, cannot transfer information contained in that record to the symbolic executor environment. The information that is stored in the PCB is, for example, the opened file descriptors and their state, the stack canary, the brk value and so on. While some information in the PCB can be retrieved using syscalls, other kinds of information available in kernel space cannot be retrieved in a similar way, like seccomp filters\footnote{Seccomp filtering provides a means for a process to specify a filter for
incoming system calls.}.

When transferring the symbolic executor state to the debugger we cannot write these values in the debugged internal process space. Additionally, tracking all the side effects at user level may be extremely challenging and add a significant overhead.

Outside the userspace context, it is possible to have full access to the operating system using an emulator like QEMU \cite{qemu} to run the target program or using a hypervisor to access kernel memory.
Furthermore, for the user, kernel debugging is not a convenient task if he needs only to debug a user-space application. An error during kernel debugging leads to an operating system reboot, possibly wasting the time spent by the analyst.

Our technique cannot be used if the analyst does not know how the execution of the debugged process until the target point affected the variables that he wants to set as symbolic. As we said before a wrong constraints selection can lead to unfeasible executions and so the analyst is limited to use this technique only when he knows the behavior of the previously executed code.

The divergence between the symbolic execution and the concrete execution is a possible error, caused, for example, by the concretization of a symbolic value read from a file to a value that is not really in that file when we try to execute the found path in the debugger. A similar issue is present in DSE (section \ref{dynamic_limit}).

\chapter{Implementation}

In this chapter, we will illustrate our {\em AngrDBG} library, a debugger-agnostic implementation of the technique described in Chapter 2 based on the binary analysis framework {\em angr}. We will present also two frontends based on that library, one for the {\em GNU Debugger}, {\em AngrGDB}, and the other for the {\em IDA Pro} debugger, {\em IDAngr}.

\section{Angr Overview}

{\em Angr} \cite{shoshitaishvili2016state} is more than just a symbolic execution engine but we will focus on this aspect.

A particular feature is that the framework is cross-architecture and cross-platform. Additionally, angr has well documented and extremely easy-to-use first-level APIs.

Angr is written in Python with some C wrappers. This may affect the performance of the engine but most of the analysis techniques are algorithmically slow and so the language performance is not very relevant.

Angr is under constant development and is made up of several modules. We will present only the ones that are relevant to our approach:

\begin{itemize}
\item {\em CLE}, a multi-format binary loader with an easy API;
\item {\em archinfo}, a collection of classes that contain architecture-specific information;
\item {\em PyVEX}, a wrapper around Valgrind's VEX IR lifter \cite{Net:phd2004}, used to make the analyses architecture-agnostic;
\item {\em Claripy}, the angr data backend, a wrapper around the Z3 solver and an interface to abstract concrete and symbolic values handling;
\end{itemize}

The symbolic execution engine is in the main package of angr, as other analyses and utilities like CFG recovery, value-set analysis, and data dependence analysis.

One of the main advantages of angr for our purpose is that everything is like a plugin. In the implementation of our technique, we have changed some low-level bits in angr without the need to fork the project.

\subsection{Fundamental Classes}

According to \cite{angrdoc} the fundamental high-level types of angr are the following:

\subsubsection{Project}

The {\em Project} is the control base in angr. Project creation invokes the CLE loader. With a Project, you are able to initialize a symbolic execution engine or dispatch an analysis.

\subsubsection{BitVec}

A {\em bitvector} is just a sequence of bits. In Angr this is the basic datatype. A bitvector can be concrete or symbolic. Performing operations with bitevectors will yield a tree of operations that are translated into constraints for the SMT solver.

\subsubsection{Loader}

The {\em CLE loader} gets the representation in a virtual address space of a program from a binary file. It solves relocations and loads also the needed libraries. CLE supports many formats and operating systems binaries, including ELF for Linux and PE for Windows.

\subsubsection{SimState}

A {\em SimState} is the representation of the program state inside the symbolic executor. Memory, registers, file system and all data that can change during the execution are in the state.
These entities are organized in plugins and are accessible using the proper field. For example, the memory plugin can be accessed with \verb|state.memory| and the plugin that handles environment interaction on POSIX-compatible systems is accessible with \verb|state.posix|.

Must-know plugins are regs and mem. With \verb|state.regs.<register name>| you can inspect the state registers and with \verb|state.mem[address].<type>| you can inspect the memory, where  \verb|<type>| is a standard type such as the integer types defined inside the {\tt stdint.h} or the string type.

\subsubsection{Simulation Manager}

The {\em simulation manager} is the primary interface to the symbolic execution engine (used for the exploration). It is the way to get the next state from the current.

The exploration can be driven with {\em find} and {\em avoid} conditions. These conditions may be addresses that must be reached in order to terminate the exploration of a path and classify it as valid or not.
They can be also functions that inspect the current state to determinate if the exploration has found a result.

\subsection{Memory Plugin}

The angr memory structure is quite complex and multi-layered. In this section, we present only a very high-level overview. 

The main goal of the memory plugin is to provide an interface for the load and store primitives for both concrete and symbolic values.

First of all, outside the memory plugin, the executable file is loaded and mapped by CLE in a so-called {\em Clemory} object. This is used by the memory plugin as a read-only memory backer to lazily load into the process address space the initial content of the memory for a binary.

The memory plugin follows a page-oriented model.
This is realized through the abstract {\em Page} class, whose default concrete implementation is based on a Python list. A Page instance indexes all objects in the memory space associated with the corresponding page.

Page objects are managed by an instance of the {\em SimPagedMemory} class.
SimPagedMemory loads pages in a lazy way based on the requests to the memory plugin. A Page content is initialized to concrete values if the requested page address is in the range of the associated Clemory object. When the page does not belong to the Clemory object (e.g. a stack page) it is left uninitialized to make a request return an unconstrained symbolic value.

The high-level class that represents the process memory space is {\em SimMemory}. The load/store requests are mapped to the paged memory keeping also a record for symbolic memory addresses. 

In VEX, the intermediate representation used by angr, registers are mapped to a separate memory, so a SimMemory is used also with registers operations.

\subsection{Simprocedures}

A {\em SimProcedure} is used to define a hook in angr. Simprocedures are mainly used with external functions hooking the associated symbol. In this way angr can skip the exploration of known library functions.

A SimProcedure is essentially a model of the function behavior written with the aim of minimizing the number of forks generated by the SimProcedure itself.

\section{AngrDBG}

{\em AngrDBG} is the library that we developed to synchronize a concrete process state with an angr state.

angr, as we said in the previous section, is composed of plugins and AngrDBG exploits this feature to modify parts of angr.

The fundamental function exposed to the user is {\em StateShot}. StateShot initializes an angr state with the current debugger context and changes the memory plugin of the state with a custom version.
The memory plugin is responsible for requesting concrete memory from the debugger when the engine wants to access a memory region mapped in the process.

A state returned by StateShot can be used like any other angr state.

The project is a global instance created lazily with \verb|load_project| that uses the same image base of the debugged process.

\subsection{Abstract Debugger}

We designed this library to be debugger independent using an abstraction layer. So AngrDBG is compatible with various tools that have Python bindings.

{\em Debugger} is the virtual class that makes this possible. Any tool that wants to synchronize angr with a specific debugger needs to implement a subclass of the Debugger class.
The global \verb|register_debugger(debugger)| routine must be invoked to tell AngrDBG to use a specified instance of a Debugger subclass as the source of data for the synchronization.

\verb|get_debugger| is used to get the currently registered instance.

The methods that must be implemented are the following:

\begin{itemize}
\item \verb|before_stateshot(self)| An event handler triggered before the synchronization setup in StateShot, just after the empty state creation;
\item \verb|after_stateshot(self, state)| An event handler triggered before the StateShot return;
\item \verb|is_active(self)| Return True if the debugger is running the target process;
\item \verb|input_file(self)| Return a Python file-like object of the target executable;
\item \verb|image_base(self)| Return the process base address;
\item \lstinline{get_<byte|word|dword|qword>(self, addr)} Read an byte|word|dword|qword from the memory as a Python int (4 distinct methods). The endianness must be concordant with the debugged process architecture;
\item \verb|get_bytes(self, addr, size)| Read a string from the memory;
\item \lstinline{put_<byte|word|dword|qword>(self, addr, value)} Write a Python int as a byte|word|dword|qword to memory (4 distinct methods). The endianness must be concordant with the debugged process architecture;
\item \verb|put_bytes(self, addr, value)| Write a string to memory;
\item \verb|get_reg(self, name)| Get a register value;
\item \verb|set_reg(self, name, value)| Set a register value;
\item \verb|step_into(self)| Call the debugger "step into" command;
\item \verb|run(self)| Run the process inside the debugger;
\item \verb|wait_ready(self)| Wait until the debugged process is ready to be inspected;
\item \verb|refresh_memory(self)| Refresh the information about the memory space inside the debugger. This is needed when the debugger uses a cache for this information;
\item \verb|seg_by_name(self, name)| Get a Segment object by the name;
\item \verb|seg_by_addr(self, name)| Get a Segment object by the address;
\item \verb|get_got(self)| Get a tuple (start address, end address) related to the GOT section when using the ELF file format;
\item \verb|get_plt(self)| Get a tuple (start address, end address) related to the PLT section when using the ELF file format;
\item \verb|resolve_name(self, name)| Resolve a symbol to its address using the name;
\end{itemize}

\subsection{Memory Synchronization Types}
\label{angrdbg_memtypes}

In AngrDBG a user can choose between four different types of memory synchronization.

The related functions are \verb|get_memory_type()| and \verb|set_memory_type(mem_type)| where \verb|mem_type| is one of the following constants:

\subsubsection{SIMPROCS\_FROM\_CLE}

The concrete memory is from the target process but, when using the ELF file format, the SimProcedures are maintained as in a regular empty angr state generated using CLE data. This is done to avoid the execution of library code but at the same time support self-modifying code. When an external function does not have a SimProcedure, the symbol is resolved and the GOT section is populated with the real address to avoid the exploration of the loader code.

The PE format is not currently supported by this type of memory synchronization.

\subsubsection{ONLY\_GOT\_FROM\_CLE}

Like the previous one, but with a SimProcedure stub that returns an unconstrained symbolic value used to model unknown imported functions.

The PE format is not currently supported by this type of memory synchronization.

\subsubsection{USE\_CLE\_MEMORY}

The segments associated with the executable file are borrowed from the CLE loader and only the segments created at runtime, like the stack, are from the concrete process. This is not accurate if the program modified some variables in .data or if it has a self-modifying code.

\subsubsection{GET\_ALL\_DISCARD\_CLE}

All the memory is from the target process, on ELFs should be used only with \verb|LD_BIND_NOW| to avoid the exploration of loader code. This should be the preferred option on Windows at the moment.

\subsection{Memory Plugin}

To synchronize the memory we extended the memory plugin to load values from the debugged process.

The replacement of the Page class is {\em DbgPage}. It is a list of bytes that works lazily. When a load request occurs, the requested data is returned if it is in the list, otherwise, if the \verb|from_dbg| attribute is present in the DbgPage object the correspondent bytes are requested to the debugger. Without the \verb|from_dbg| attribute, so when the page does not belong to the concrete address space, an unconstrained symbolic value is returned.

{\em SimDbgMemory} is the correspondent class to SimPagedMemory. We patched the page initialization method. If the memory type is \verb|USE_CLE_MEMORY| and the requested page is in the Clemory backer it loads the page from the backer like angr usually does. In the other situations the page is initialized as empty (remember that DbgPage works lazily) and the permissions are got from the debugger when possible.

{\em SimSymbolicDbgMemory} is the high-level class, simply a version of SimMemory that works with SimDbgMemory instead of SimPagedMemory.

In the StateShot routine, the state is created specifying SimSymbolicDbgMemory as the memory plugin.

\subsection{Internal process state synchronization}

We implemented the retrieval of the relevant process related values in the operating system executing syscalls on-the-fly.

The method is the following:

\begin{itemize}
\item Save the context of the concrete process and the first bytes of the executable segment;
\item Write the syscall instruction (int 80h on x86) at the beginning of the executable segment;
\item Set input registers and the program counter to the syscall instruction address;
\item Execute the syscall with a single step in the debugger;
\item Retrieve the results;
\item Restore the context and the previously patched bytes;
\end{itemize}


The current implementation supports the retrieving of the BRK\footnote{The top of the Heap segment} value.

File and network operations return by default symbolic values. Reading from a file descriptor gives uninitialized symbolic memory. A custom behavior (e.g. load concrete data from a file in the disk) can be forced by creating an \verb|angr.SimFile| object and linking it to the initial state, as explained in \cite{angr_file_doc}.

Segment registers are missing from the synchronization. At the moment all memory read using a segment register (like the stack canary) is symbolic.

\section{Using AngrDBG API}

AngrDBG comes with a wrapper class around StateShot, {\em StateManager}, that allows the user to easily create a state from the debugger state and manage the symbolic values creation and concretization. This class is also responsible to keep track of the symbolic inputs and inject them into the debugger after the exploration.

The exposed methods are the following:

\begin{itemize}
\item \verb|sim(self, key, size=None)| Set a value as symbolic. Key can be an address (int) or a register name (string). If the size parameter is not specified it is the default register size when the key is a register or the size of a pointer in the debugged program architecture when the key is an address.
\item \verb|__getitem__(self, key)| The get operator. If the key is a register name return the associated value. If the key is an address access to the state memory in the same way of angr with \verb|state.mem|.
\item \verb|__setitem__(self, key, value)| The set operator. If the key is a register name overwrite the register content. If the key is an address write to the state memory in the same way of angr with \verb|state.mem|.
\item \verb|simulation_manager(self)| Load the global project and generate a simulation manager based on the current state.
\item \verb|to_dbg(self, found_state)| Concretize the corresponding symbolic values in \verb|found_state| and write that values in the debugger. \verb|found_state| can be an angr state or another StateManager instance.
\item \verb|concretize(self, found_state)| Like \verb|to_dbg| but return the concretized value in a dictionary instead of writing the values to the debugger.
\end{itemize}

\subsection{Remote Server}

Angr explorations require a huge amount of RAM. To allow the usage of angr in a powerful hardware setup and improve the exploration time AngrDBG and the debugger can run in different machines.

The command \verb|python -m angrdbg| starts an rpyc \cite{rpyc} based server. The default host is localhost in order not to expose an unsecure connection. The user must do SSH port forwarding to set up a secure connection.

The angrdbg server waits for two rpyc connections and after that, it opens an IPython \cite{ipython} kernel in the current TTY. The first connection is used to serve data to the IPython shell, so the roles of client-server are swapped in this case. The second is used to wait for remote procedure calls from the client.

The two connection lives in different processes to avoid race conditions.


\section{AngrGDB}

{\em AngrGDB} is in its core an implementation of the Debugger class of AngrDBG using the GDB Python bindings.

It works also with remote debugging sessions so you can, for example, attach GDB to QEMU and perform symbolic execution using AngrGDB.

The methods \cite{gdb_py_doc} that are invoked to retrieve information are gdb.execute, used to execute a GDB command (e.g. "info address"), \verb|write_memory| and \verb|read_memory|.

\subsection{Using AngrGDB}

To use AngrGDB you can simply open a Python shell under GDB (with the "pi" command) and use the AngrDBG API.

In addition, we defined some custom GDB commands to easily explore code using find and avoid targets.

The commands are the following:

\begin{code}
angrgdb sim <register name> [size]
angrgdb sim <expression> [size]
\end{code}
Set a memory/register as symbolic

\begin{code}
angrgdb list
\end{code}
List all items that you setted as symbolic

\begin{code}
angrgdb find <address0> <address1> ... <addressN>
\end{code}
Set the list of find targets

\begin{code}
angrgdb avoid <address0> <address1> ... <addressN>
\end{code}
Set the list of avoid targets

\begin{code}
angrgdb reset
\end{code}
Reset the context (symbolic values and targets)

\begin{code}
angrgdb run
\end{code}
Generate a state from the debugger state and run the exploration

\begin{code}
angrgdb shell
\end{code}
Open an IPython shell with a StateManager instance created from the current GDB state called "sm"

\section{IDAngr}

{\em IDAngr} is not simply an AngrDBG frontend library for the IDA Pro \cite{ida} debugger but it has also a graphical interface in Qt integrated directly in IDA.

It can be loaded using the "load script" command in the File menu or installed as a plugin and invoked using the Ctrl-Alt-I keyboard shortcut.

IDAngr implements also the remote AngrDBG protocol and can be attached to a remote AngrDBG server, as explained in section 3.3.1.

Before to use the AngrDBG API the library must be initialized.

The first method that must be called is \verb|idangr.init|. Without specifying arguments, the AngrDBG and Angr modules will be imported using the regular python import statement and the exploration will be performed in the same machine running IDA Pro. If the host and port arguments are present the selected remote AngrDBG server will be connected to IDAngr.

\subsection{GUI Overview}

The initialization process can be done also using the GUI. When the plugin is called for the first time a popup window asks the user if he wants to initialize a local or a remote AngrDBG instance.

The graphical components added to IDA are two: a context menu entry in the menu shown when clicking the right button of the mouse and a panel inserted alongside the IDA View.

\begin{figure}
    \centering
    \parbox{5cm}{
        \caption{The context menu.}
        \includegraphics[width=5cm]{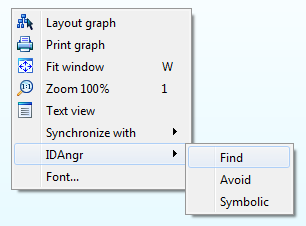}
        \label{fig:ctx_menu}
    }
    \qquad
    \begin{minipage}{5cm}
    \caption{The Add symbolic memory prompt.}
    \includegraphics[width=5cm]{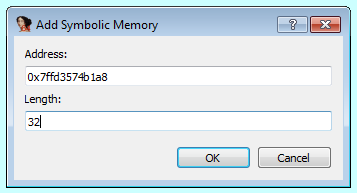}
    \label{fig:add_symbolic}
    \end{minipage}
\end{figure}

In the context menu, Figure \ref{fig:ctx_menu}, you can perform the following actions:

\begin{itemize}
\item {\em Find}: add the current selected address to the simulation manager's find targets;
\item {\em Avoid}: add the current selected address to the simulation manager's avoid targets;
\item {\em Symbolic}: symbolize a memory region starting from the current selected address and specifying its size, Figure \ref{fig:add_symbolic};
\end{itemize}

\begin{figure}[H]
  \caption{The symbolic values and the find/avoid lists in the panel.}
  \centering
  \includegraphics[width=1\textwidth]{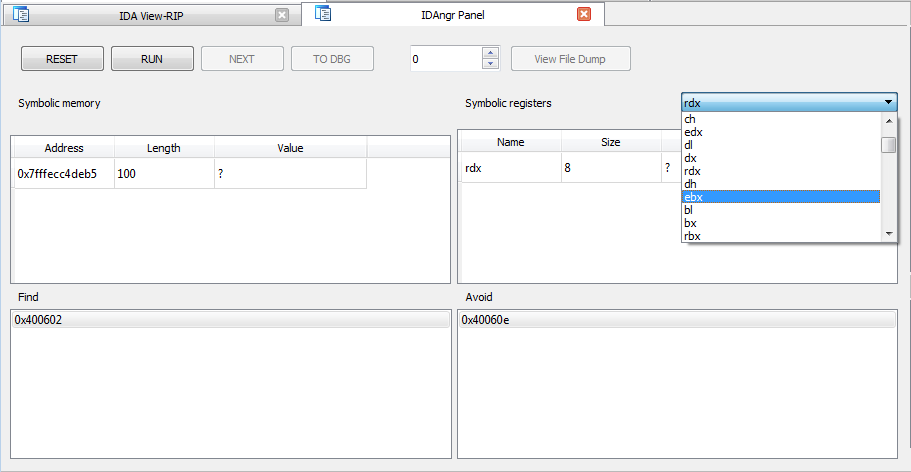}
  \label{fig:panel1}
\end{figure}

The IDAngr panel is the core graphical component of the plugin. In the panel, you can view the values selected using the context menu. You can also set symbolic registers, as in Figure \ref{fig:panel1}.

\begin{figure}[H]
  \caption{Setting preconditions on a symbolic value.}
  \centering
  \includegraphics[width=0.8\textwidth]{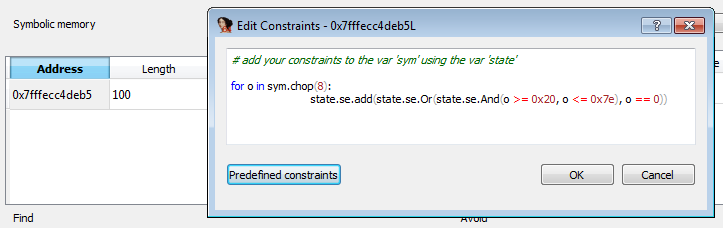}
  \label{fig:precond}
\end{figure}

By right-clicking on a row of the symbolic memory or registers views you can set the preconditions using python (figure \ref{fig:precond}), jump to its address in the IDA view or remove it.

By right-clicking on a row of the find or avoid view you can jump to its value or delete it.

\begin{figure}[H]
  \caption{The exploration prompt.}
  \centering
  \includegraphics[width=0.6\textwidth]{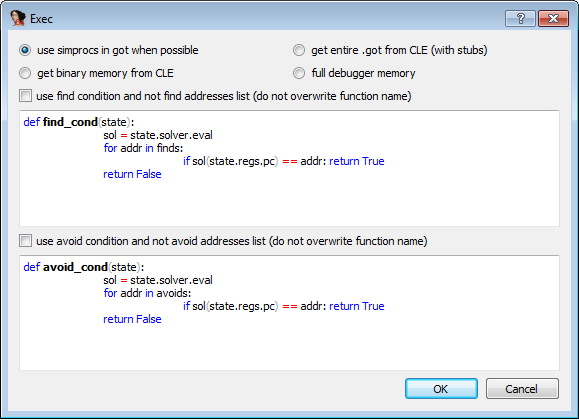}
  \label{fig:run}
\end{figure}

The buttons at the top of the panel are the following:

\begin{itemize}
\item {\em RESET}: Delete all symbolic values and find/avoid targets
\item {\em RUN}: Show the exploration prompt, Figure \ref{fig:run}, in order to start the symbolic exploration
\item {\em NEXT}: Show the exploration prompt, Figure \ref{fig:run}, in order to reach another valid state after a previous exploration
\item {\em TO DBG}: Use a valid found state to concretize all symbolic values listed in the panel and inject them in the debugger
\item {\em View File Dump}: Use a valid found state to dump the content of a file using its file descriptor
\end{itemize}

The exploration prompt is the last step left to the user before the symbolic exploration. This window allows you to set which memory type use (based on Section \ref{angrdbg_memtypes}) and set a python function as find or avoid condition instead of using the list of targets in the panel.

The correspondent types of memory synchronization:

\begin{itemize}
\item \verb|SIMPROCS_FROM_CLE|: use simprocs in got when possible
\item \verb|ONLY_GOT_FROM_CLE|: get entire .got from CLE (with stubs)
\item \verb|USE_CLE_MEMORY|: get binary memory from CLE
\item \verb|GET_ALL_DISCARD_CLE|: full debugger memory
\end{itemize}

Hooks are at the moment not supported in the graphical interface, but they can be used from IDAPython as regular angr hooks. Simply use \verb|load_project| to get the current project and apply hooks on it:

\begin{py_code}
p = angrdbg.load_project()
@p.hook(0xdeadbeef)
def hook(state):
    print "I'm at", state.regs.rip
\end{py_code}


\chapter{Case Study}

In this chapter, we will carry out a small reversing engineering task using IDAngr to show a usage example of the tool.

\section{Task description}

The task is reversing a custom hash function in an ELF binary for Linux x86\_64. The name of the binary is sftp \footnote{The reader can download the binary here: \url{https://github.com/andreafioraldi/bsc-thesis/raw/master/case_study/sftp}}, a binary exploitation challenge from Google CTF 2018 \cite{gctf} that requires a password to use the program. 

That password hash is hardcoded and the interesting code is the function at address 0x13F0, called by the main.

The rest of the binary is an in-RAM filesystem with a buggy allocator. The challenge was to exploit that bug but we will discuss only the part related to the password discovery.

The output of the decompiler for the function is the following:

\begin{cpp_code}
signed __int64 sub_13F0()
{
  char *v0; // rbx
  __int64 v1; // rdx
  signed __int64 result; // rax
  int v3; // eax
  int v4; // edx
  char v5; // [rsp+0h] [rbp-28h]
  char v6; // [rsp+1h] [rbp-27h]
  char v7; // [rsp+2h] [rbp-26h]
  char v8; // [rsp+3h] [rbp-25h]
  unsigned __int64 v9; // [rsp+18h] [rbp-10h]

  v9 = __readfsqword(0x28u);
  v0 = &v5;
  __printf_chk(1LL, "The authenticity of host '
  puts("ECDSA key fingerprint is SHA256:+d+dnKGLreinYcA8EogcgjSF3yhvEBL+6twxEc04ZPq.");
  __printf_chk(1LL, "Are you sure you want to continue connecting (yes/no)? ", v1);
  if ( !(unsigned int)__isoc99_scanf("
    return 0LL;
  if ( v5 != 'y' )
    return 0LL;
  if ( v6 != 'e' )
    return 0LL;
  if ( v7 != 's' )
    return 0LL;
  if ( v8 )
    return 0LL;
  __printf_chk(1LL, "Warning: Permanently added '
  __printf_chk(1LL, "
  if ( !(unsigned int)__isoc99_scanf("
    return 0LL;
  v3 = _IO_getc(stdin);
  LOWORD(v3) = v5;
  if ( !v5 )
    return 0LL;
  v4 = 21527;
  do                                            // hashing
  {
    v3 ^= v4;
    ++v0;
    v4 = 2 * v3;
    LOWORD(v3) = *v0;
  }
  while ( (_BYTE)v3 );
  result = 1LL;
  if ( (_WORD)v4 != -29190 )
    return 0LL;
  return result;
}
\end{cpp_code}

\section{Solution}

To bypass the authentication we will use IDAngr.
We are using IDA Pro on Windows and so we start also a remote AngrDBG server on a Linux machine.

\begin{figure}[H]
  \centering
  \includegraphics[width=1\textwidth]{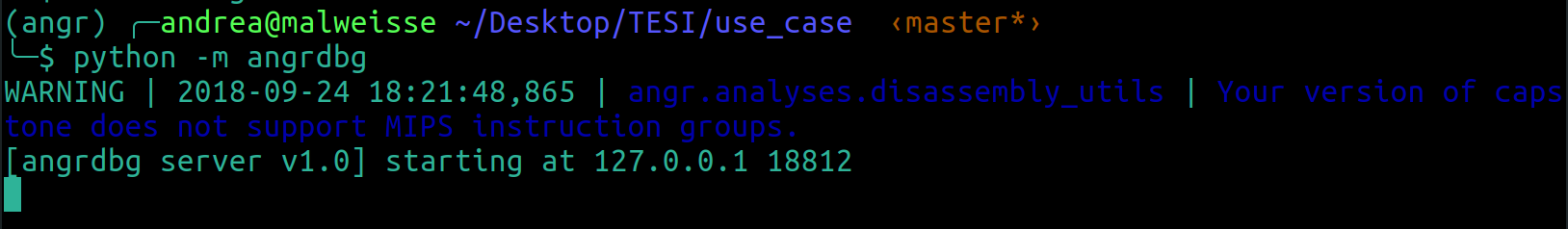}
  \label{fig:u1}
\end{figure}

The first step in IDA is to start the debugger and set a breakpoint at line 34 of the function.
This line is just after a scanf call. The inserted string is the password.

As you can see at line 47 the final hash must be -29190, 0x8dfa in hexadecimal.

Running the debugger with F9 the debugged program asks the user to insert the string yes on the standard input.

After that, the program is waiting for the input again using scanf.
Looking at the scanf format \verb|"%15s"| we can know that the password length maximum value is 15, so we insert \verb|aaaaaaaaaaaaaaa| as dummy data.

\begin{figure}[H]
  \centering
  \includegraphics[width=1\textwidth]{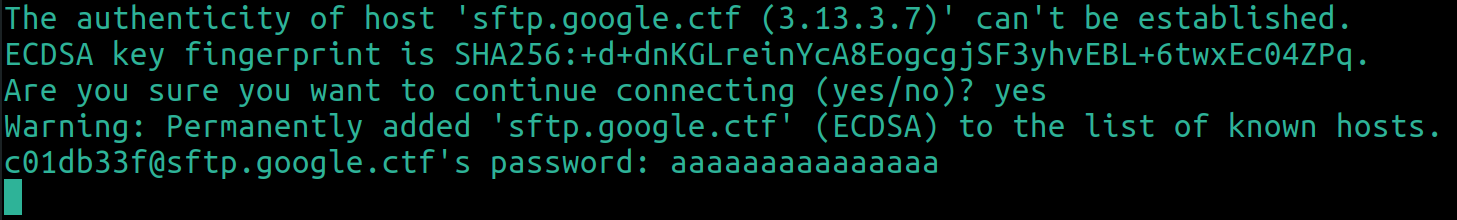}
  \label{fig:u2}
\end{figure}

Now the execution has just hit our breakpoint.
At this point, we must invoke the plugin with Ctrl-Alt-I.

The initialization popup asks if we want to use a remote instance of AngrDBG and we insert the correct IP address and port of our Linux machine running the server.

\begin{figure}[H]
  \centering
  \includegraphics[width=0.6\textwidth]{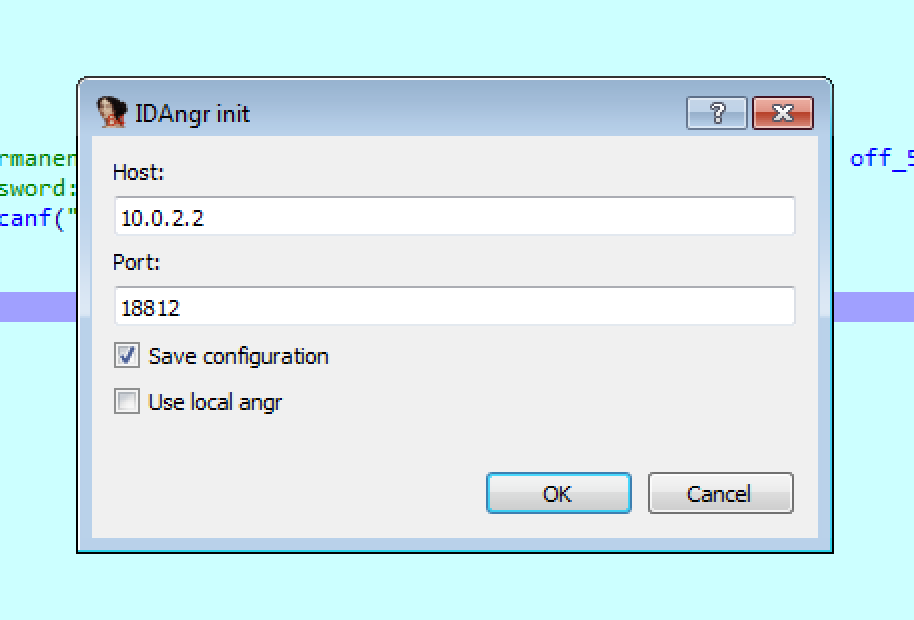}
  \label{fig:u3}
\end{figure}

Firstly we symbolize the input using the context menu:

\begin{figure}[H]
  \centering
  \includegraphics[width=0.6\textwidth]{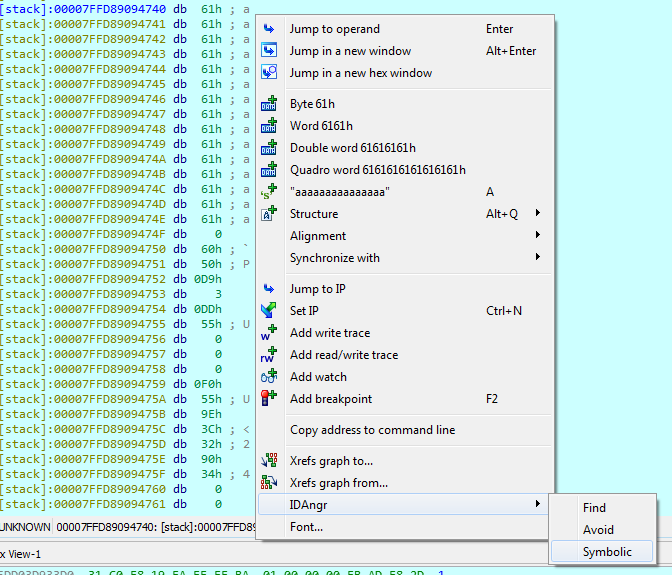}
  \label{fig:u4}
\end{figure}

In the popup, we set the length of the symbolic value, 15:

\begin{figure}[H]
  \centering
  \includegraphics[width=0.8\textwidth]{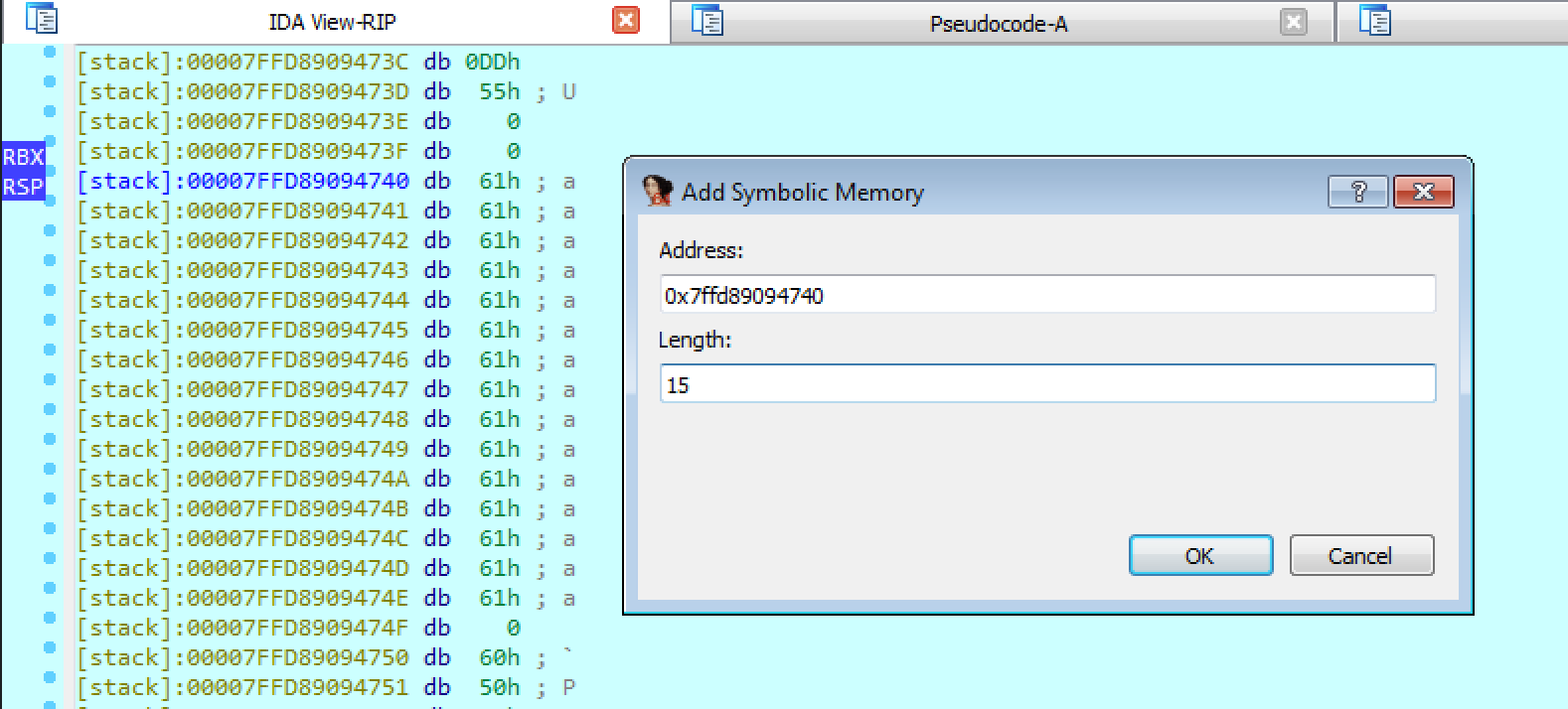}
  \label{fig:u5}
\end{figure} 

We want also to add pre-constraints to the symbolic input. The password must be composed of printable characters and must not have spaces in order to be compatible with scanf.

By right-clicking on the symbolic value in the panel and selecting \verb|add constraints| we can insert a python snippet and add these constraints:

\begin{figure}[H]
  \centering
  \includegraphics[width=0.4\textwidth]{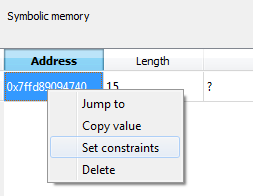}
  \label{fig:u6}
\end{figure}

\begin{figure}[H]
  \centering
  \includegraphics[width=0.8\textwidth]{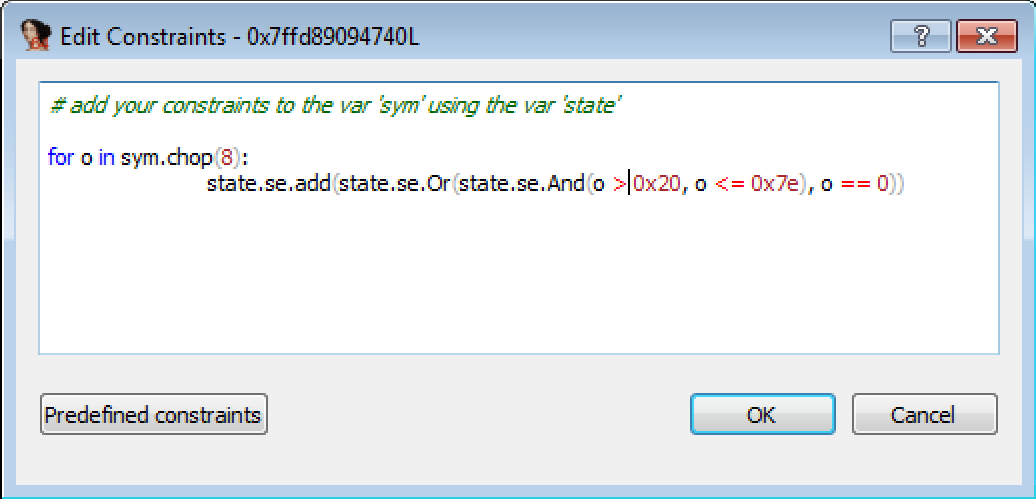}
  \label{fig:u7}
\end{figure} 

The next step is to select the find and avoid targets. The interesting branch is the hash value comparison at line 47. We want to avoid line 48 and execute line 49. 

In the graph view, the branch is the following. Line 48 is the basic block on the left, line 49 is the block on the right.

\begin{figure}[H]
  \centering
  \includegraphics[width=0.8\textwidth]{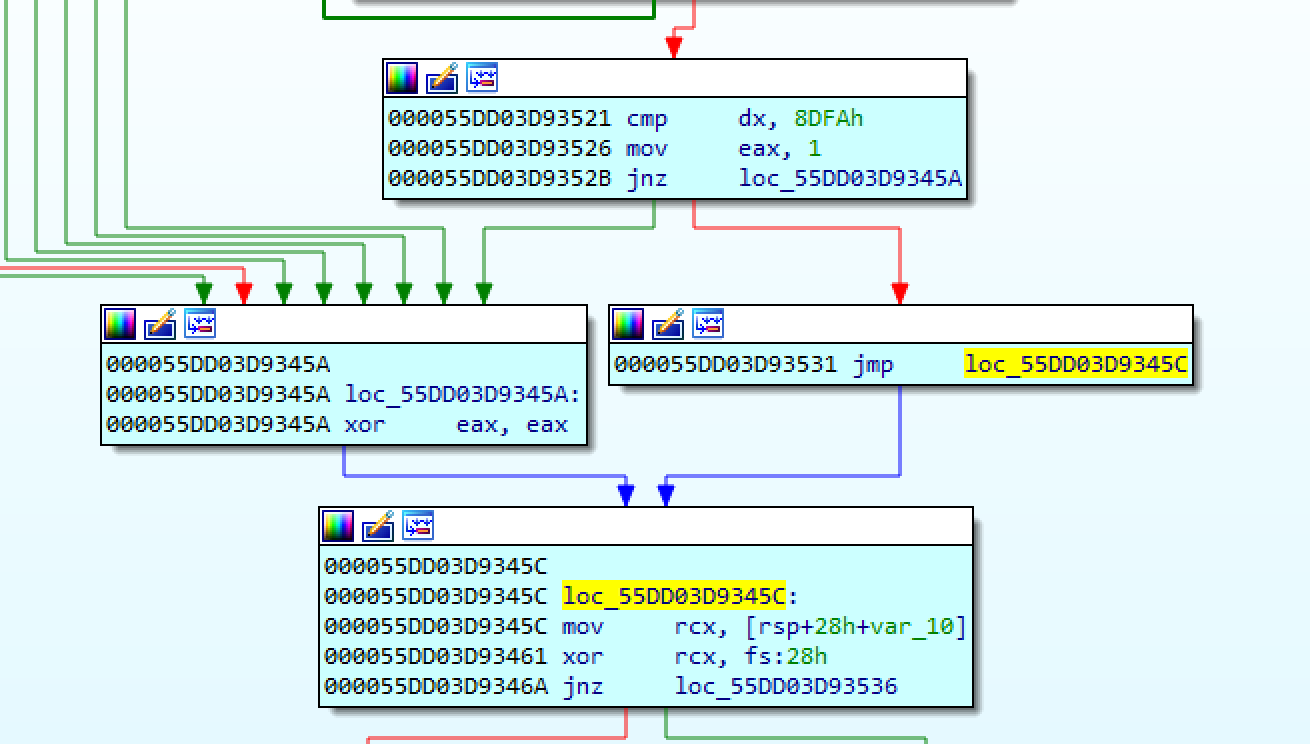}
  \label{fig:u8}
\end{figure} 

Using the context menu we set the instruction addresses corresponding to the two lines of code as avoid and find targets, respectively:

\begin{figure}[H]
  \centering
  \includegraphics[width=0.8\textwidth]{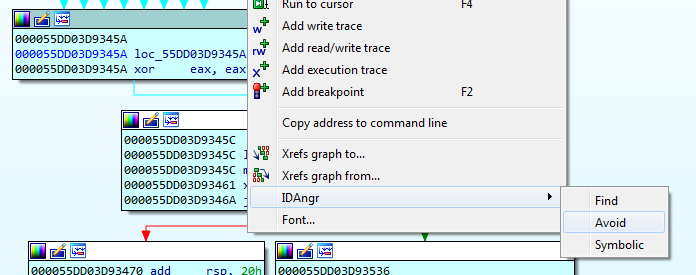}
  \label{fig:u9}
\end{figure} 

We can now run the exploration clicking on the RUN button to show the exploration prompt. In this prompt we leave as selected the default memory type, \verb|use simprocs in got when possible|. We are not exploring external code so all the memory types have the same behavior.

\begin{figure}[H]
  \centering
  \includegraphics[width=0.9\textwidth]{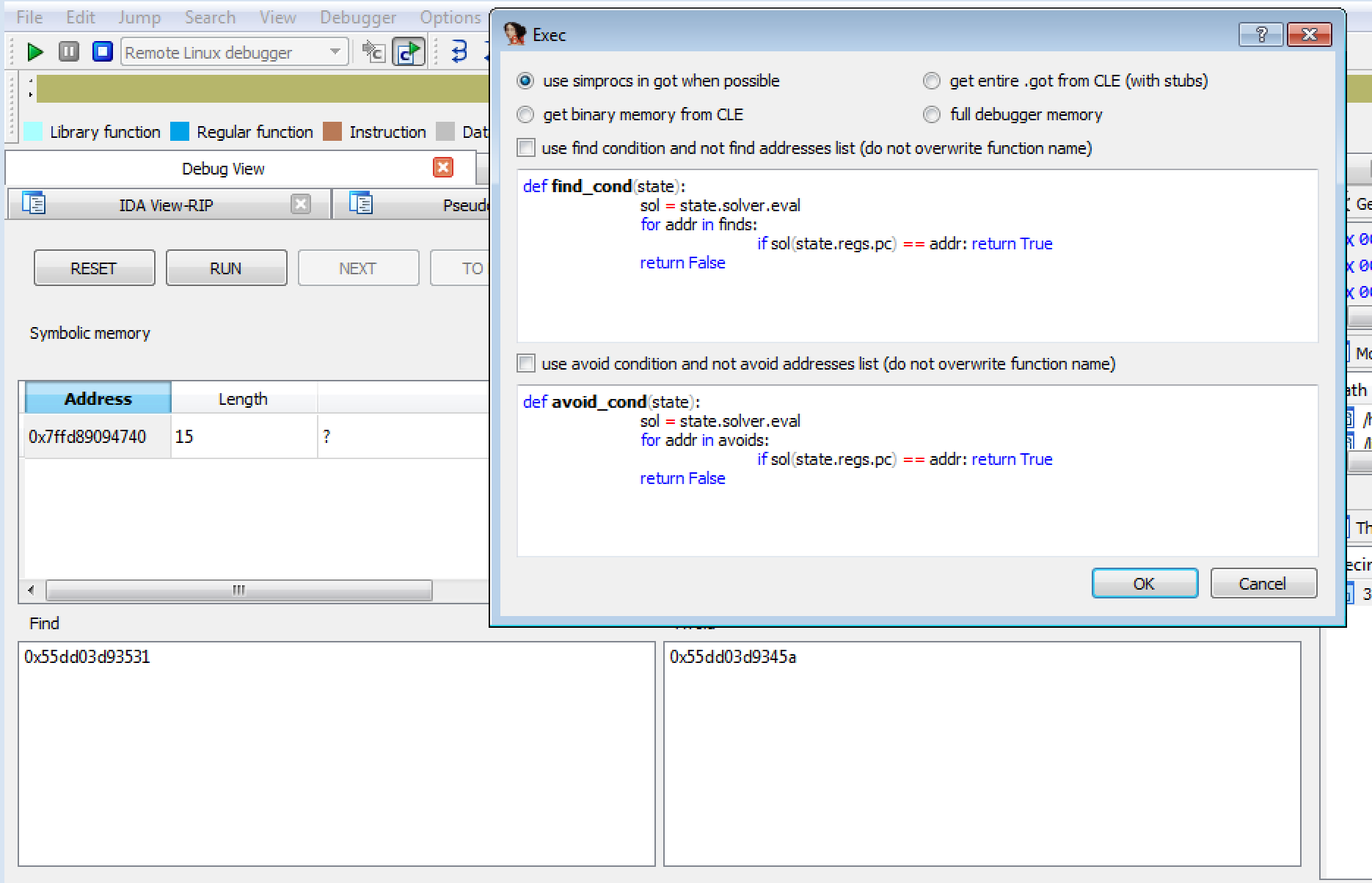}
  \label{fig:u10}
\end{figure} 

We click on OK to start the exploration.
After some time we get a valid concretized value of the password in the panel. 

\begin{figure}[H]
  \centering
  \includegraphics[width=1\textwidth]{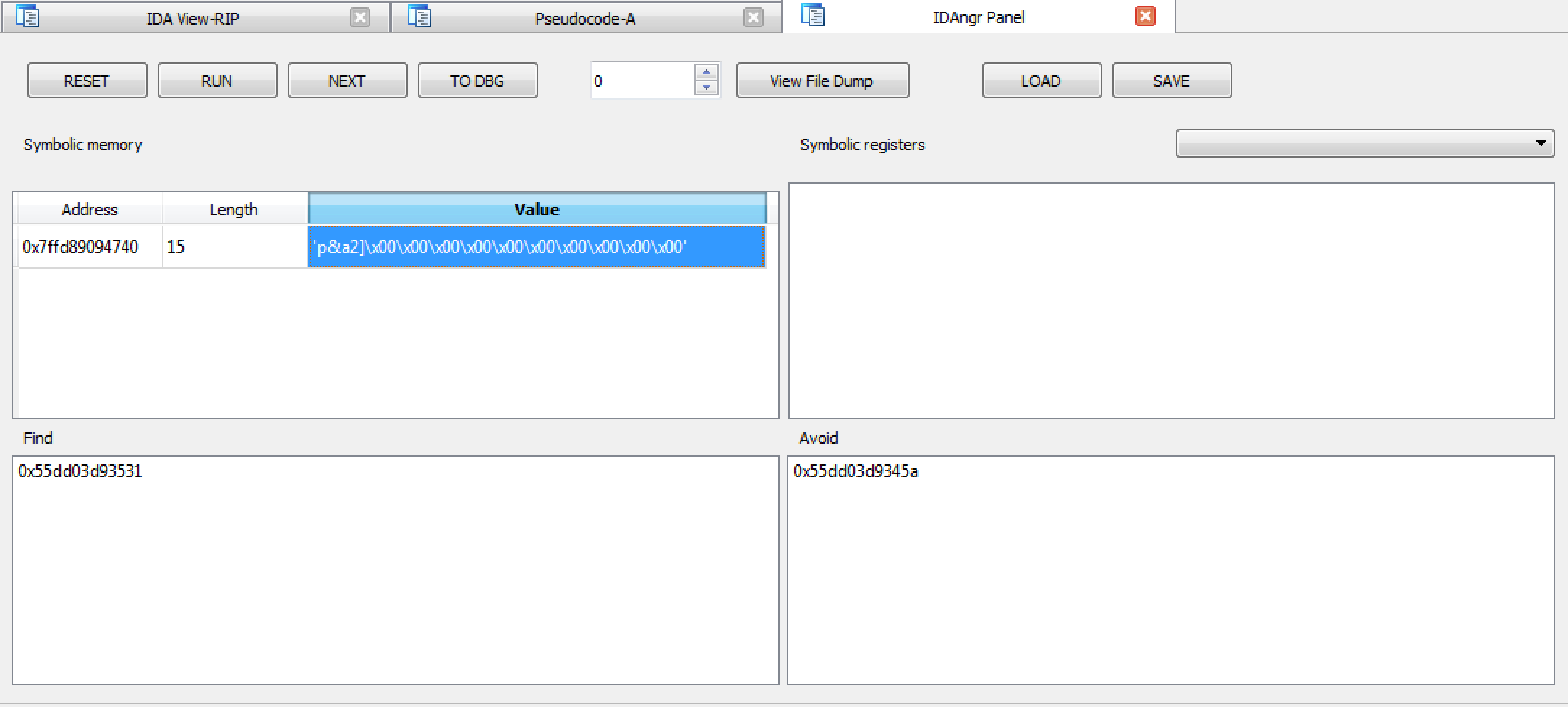}
  \label{fig:u11}
\end{figure}

This is our solution. We can now use the "TO DBG" button to inject this value in the debugger memory:

\begin{figure}[H]
  \centering
  \includegraphics[width=0.4\textwidth]{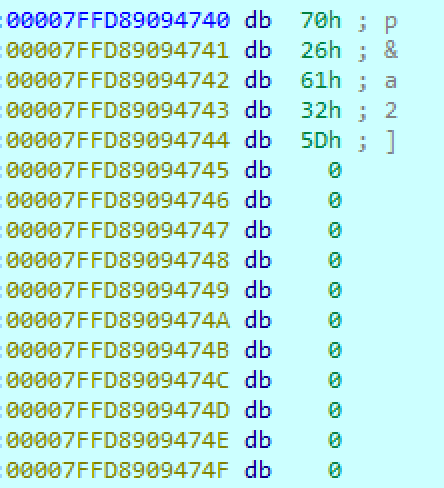}
  \label{fig:u12}
\end{figure}

By stepping with the debugger until the end of the function we reach line 49 rather than line 48. The authentication is successful.

\begin{figure}[H]
  \centering
  \includegraphics[width=0.4\textwidth]{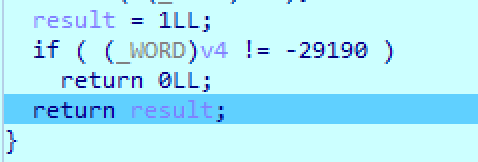}
  \label{fig:u13}
\end{figure}

\section{Discussion}

This simple use case, in our opinion, shows how user-friendly the IDAngr graphical interface is.
The user interaction is a crucial feature of the technique. Reverse engineering is and will be a manual task and having a comfortable interface make a difference.
Of course, state synchronization is not limited to the exploration of simple pieces of code like the one in the example, but this Google CTF challenge is representative of the workflow that a reverser can carry out assisted by our tool.

\chapter{Conclusion}

Reverse engineering a software is a fundamental task for improving the security of systems. In the last years, its perceived importance is increasing exponentially due to the enormous number of malware discovered every day.

Research on new methods and technologies to improve this process is crucial. Our contribution is a solution that can speed up the debugging of complex software using symbolic execution.

As current limitations, the tools and libraries that we developed are effective but far from to be complete. Windows support is poor and environment interaction must be improved.

State synchronization is already present in literature \cite{muench:bar18} but with a partial approach (state transfer in one direction) and without the focus on user interaction. Our solution, instead, is heavily based on the interaction with the user with the aim to integrate symbolic execution into a debugger using a comfortable interface. We built a tool with a graphical interface to simplify as much as possible the symbolic executor setup by the analyst.

\section{Future work}


A future direction we plan to explore is to develop an AngrDBG frontend for a DBI framework. We considered to use {\em Frida} \cite{Frida} or {\em QBDI} \cite{qbdi}. Frida can easily instrument mobile applications and so perform symbolic execution on them can be useful for a mobile security researcher. QBDI, instead, works well on x86\_64 and it has python bindings so that AngrDBG can run directly in the instrumented process without too much effort in developing the frontend.

The synchronization of the environment and simprocedures on Windows are also features that will be added.

\section{Final Words}

I want to thank all my family, friends and colleagues for the support and Prof. Camil Demetrescu, Dr. Emilio Coppa and Dr. Daniele Cono D'Elia for their patience and for their support in this project.

\vfill

{\em A little green bug has come to get squashed}

Broly to Piccolo

\newpage
\phantomsection

\bibliography{ms} 
\bibliographystyle{ieeetr}

\end{document}